\documentclass[prb,showpacs,twocolumn,aps,a4paper]{revtex4}
\usepackage{amsmath}
\usepackage{graphicx}
\usepackage{latexsym}
\usepackage{amsfonts}
\usepackage{amssymb}
\usepackage{verbatim}
\DeclareGraphicsExtensions{.pdf,.gif,.jpg}

\newcommand{\be}{\begin{equation}}
\newcommand{\ee}{\end{equation}}
\newcommand{\beq}{\begin{eqnarray}}
\newcommand{\eeq}{\end{eqnarray}}

\begin{document}

\def\bbe{\mbox{\boldmath $e$}}
\def\bbf{\mbox{\boldmath $f$}}
\def\bg{\mbox{\boldmath $g$}}
\def\bh{\mbox{\boldmath $h$}}
\def\bj{\mbox{\boldmath $j$}}
\def\bq{\mbox{\boldmath $q$}}
\def\bp{\mbox{\boldmath $p$}}
\def\br{\mbox{\boldmath $r$}}

\def\bone{\mbox{\boldmath $1$}}

\def\dr{{\rm d}}

\def\tb{\bar{t}}
\def\zb{\bar{z}}

\def\tgb{\bar{\tau}}

\def\bC{\mbox{\boldmath $C$}}
\def\bG{\mbox{\boldmath $G$}}
\def\bH{\mbox{\boldmath $H$}}
\def\bK{\mbox{\boldmath $K$}}
\def\bM{\mbox{\boldmath $M$}}
\def\bN{\mbox{\boldmath $N$}}
\def\bO{\mbox{\boldmath $O$}}
\def\bQ{\mbox{\boldmath $Q$}}
\def\bR{\mbox{\boldmath $R$}}
\def\bS{\mbox{\boldmath $S$}}
\def\bT{\mbox{\boldmath $T$}}
\def\bU{\mbox{\boldmath $U$}}
\def\bV{\mbox{\boldmath $V$}}
\def\bZ{\mbox{\boldmath $Z$}}

\def\bcalS{\mbox{\boldmath $\mathcal{S}$}}
\def\bcalG{\mbox{\boldmath $\mathcal{G}$}}
\def\bcalE{\mbox{\boldmath $\mathcal{E}$}}

\def\bgG{\mbox{\boldmath $\Gamma$}}
\def\bgL{\mbox{\boldmath $\Lambda$}}
\def\bgS{\mbox{\boldmath $\Sigma$}}

\def\bgr{\mbox{\boldmath $\rho$}}

\def\a{\alpha}
\def\b{\beta}
\def\g{\gamma}
\def\G{\Gamma}
\def\d{\delta}
\def\D{\Delta}
\def\e{\epsilon}
\def\ve{\varepsilon}
\def\z{\zeta}
\def\h{\eta}
\def\th{\theta}
\def\k{\kappa}
\def\l{\lambda}
\def\L{\Lambda}
\def\m{\mu}
\def\n{\nu}
\def\x{\xi}
\def\X{\Xi}
\def\p{\pi}
\def\P{\Pi}
\def\r{\rho}
\def\s{\sigma}
\def\S{\Sigma}
\def\t{\tau}
\def\f{\phi}
\def\vf{\varphi}
\def\F{\Phi}
\def\c{\chi}
\def\w{\omega}
\def\W{\Omega}
\def\Q{\Psi}
\def\q{\psi}

\def\ua{\uparrow}
\def\da{\downarrow}
\def\de{\partial}
\def\inf{\infty}
\def\ra{\rightarrow}
\def\lra{\leftrightarrow}
\def\bra{\langle}
\def\ket{\rangle}
\def\grad{\mbox{\boldmath $\nabla$}}
\def\Tr{{\rm Tr}}
\def\Re{{\rm Re}}
\def\Im{{\rm Im}}

\title{ Production, storage and release of spin currents in quantum circuits   }


\author{Michele Cini}

 \affiliation{ Dipartimento di Fisica, Universit\`{a}
di Roma Tor Vergata, Via della Ricerca Scientifica 1, I-00133 Rome,
Italy, and Istituto Nazionale di Fisica Nucleare - Laboratori
Nazionali di Frascati, Via E. Fermi 40, 00044 Frascati, Italy.}

\begin{abstract}
  Quantum rings connected to  ballistic circuits  couple  strongly to external magnetic fields if the connection is not symmetric. By analytical theory and  computer simulation I show that properly connected rings can be used to pump currents in the wires giving raise to a number of interesting new phenomena. One can  pump spin polarized currents into the wires by using rotating magnetic fields or letting the ring rotate around the wire. This method  works without any  need for  the spin-orbit interaction, and without stringent requirements about the conduction band   filling.  On the other hand, another method works at half filling using a  time-dependent magnetic field in the plane of the (fixed) ring. This  can be used to pump a pure spin current, excited by the the spin-orbit interaction in the ring.  One can use magnetizable bodies  as storage units to concentrate and save the magnetization  in much the same way as capacitors store electric charge. The polarization obtained in this way can then  be   used on command to produce spin currents in a wire. These currents show interesting oscillations while the storage units exchange their polarizations  and the intensity and amplitude of the oscillations can controlled by tuning the conductance of the wire used to connect the units.
\end{abstract}


\maketitle







\section{Introduction}

In the past 15  years or so, there has been in the literature a growing interest in the relation between magnetic and transport properties of materials\cite{amico}; on the other hand, there have been several works  on the  persistent as well as transient
currents in quantum rings  threaded by a magnetic flux, with a promising outlook in the quest for new device applications in spintronics, memory devices, optoelectronics, quantum pumping, and quantum information processing \cite{devices1, devices2, devices3, devices4}. Aharonov-Bohm-type thermopower oscillations of a quantum dot embedded in a ring for the case when the interaction between electrons can be neglected, were investigated in the literature, showing it to be strongly flux- and experimental geometry- dependent.\cite{saro1} Spin interference modulations of the conductance\cite{frustaglia} and Berry phase effects on the magnetoresistance\cite{nagasawa} have been studied.

The present paper is devoted to the magnetic properties of quantum rings linked to ballistic circuits. If the connection is asymmetric, a current in the external circuit selects a chirality in the ring and produces a magnetic moment; as we shall see, the reverse is also true, namely, a chiral current in the ring can pump charge in the wire.  The current excited in the wires by a local action in the ring is called pumping and is itself a purely quantum phenomenon.
By the same token, a symmetrically connected quantum ring inserted in a circuit cannot choose a chirality and has zero magnetic moment when a current flows through it. A symmetric connection  is unfavorable for quantum pumping. This is why a maximally asymmetrical connection is relevant in this respect.  I call this geometry a laterally connected ring (see Figure 1 a)). \\
 Compelling physical arguments based on thought experiments \cite{ciniperfettostefanucci}   suggest that the magnetic moment  is {\em not} obtained by substituting the quantum current in the classical formula\cite{Jackson}. The basic reason is that any measurement of the magnetic moment of the ring requires the measurement of a force exerted on the ring by a probe flux which according to Quantum Mechanics has an influence on the current. Actually, rephrasing the conclusions of Ref.\cite{ciniperfettostefanucci}, the magnetic moment  is given by the operator
 \be
 \hat{M}(\phi)=\frac{\de H^G_{ring}}{\de \phi}+\Delta^{\dagger} H^G_{ring}+ H^G_{ring}\Delta,
 \ee
where $\phi$ is the probe magnetic flux through the ring, $H^G_{ring}$ is the grand-canonical ring Hamiltonian referenced to the equilibrium chemical potential of the system and $\Delta=\frac{\de}{\de \phi}$. As a consequence, when the circuit is biased by a small potential difference $V_{bias}$  and a small current flows, $\langle \hat{M}\rangle $ is found\cite{ciniperfettostefanucci} to go with $V_{bias}^3$ while of course classically one expects $M \sim V_{bias}$ in a linear circuit. Indeed  at small $V_{bias}$ the quantum effects favor a laminar current which is not coupled to the magnetic field.\\

 A complementary way to study those quantum effects is\cite{ciniperfetto} to consider the reversed situation when $V_{bias}=0$ and  it is the interaction of the ring with a magnetic field that produces a current in the external circuit. Romeo and Citro, in a very interesting paper, have discussed memory and pumping effects in rings coupled to parasitic nonlinear dynamics\cite{pump4}. Laterally connected rings have an intrinsic non-linear dynamics\cite{ciniperfettostefanucci};
  in \cite{torino,cibe1,cibe2} it was shown that they have  peculiar properties for quantum pumping;  besides, they can be used to  pump spin, rather than just charge. In principle Quantum Mechanics allows us to build a device to achieve  that in more than one way.In the present paper I present new data to further clarify these purely quantum  properties. These findings raise new questions.   Can we build a ring  device that can produce a pure spin current, without any charge current associated to it?  Can we transfer magnetization between two distant bodies directly through a lead, without moving any charges? Can we store a spin current as we do with charge currents in batteries? In the following I show that these questions  have a positive answer.\\

   Such theoretical ideas, once realized in a practical device, would  offer new strategies to attack the problems connected to Spintronic applications.

\section{Geometry and dynamics of spin current generation}

In this Section I introduce a master Hamiltonian $H_{prod}$ which, by a proper choice of parameters, allows to study 3 different mechanisms for pumping   (partially or totally polarized) current in the absence of an external bias. Below,  $H_{prod}$ will   be complemented with other terms to allow for spin current storage and release. All three methods use a laterally connected ring.
I focus on maximally asymmetric geometries, i.e. laterally connected rings, such that the external circuit  is tangential to the ring. Fig.1 a)   shows   the geometry for a ring of N=8 sites; the side length is of order of 1 Angstrom.
\begin{figure}[]
\includegraphics*[width=.35 \textwidth]{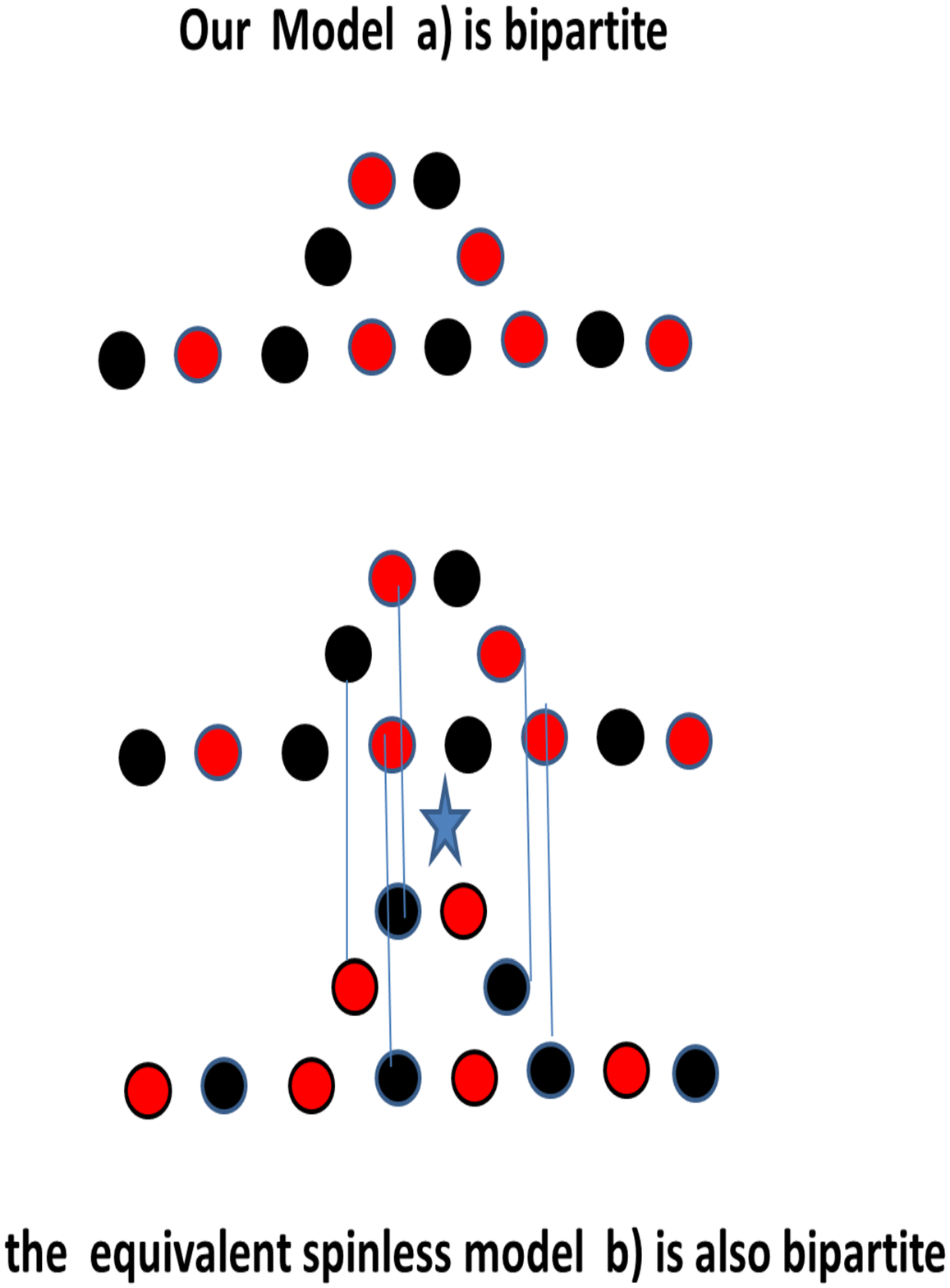}
\caption{Lateral  connection of  $N_{ring}=6$  ring to wires. (Any even $N_{ring}>2$  can be considered.) The circles  represent the sites; those in the hexagon feel the spin-orbit interaction and the magnetic field. In panel a) the geometry of the device is shown. The alternating colors visualize the bipartite property of the associated graph.The time dependent external magnetic field  has a component $B_{\|}(t)$ in the plane of the ring which couples to spin  and a component $B_{\lfloor}(t)$ which produces a flux  $\phi(t)=B_{\lfloor} S$  and also interacts with spins in the octagon; all sites are connected horizontally to the first neighbors by spin-diagonal matrix elements. In b) the adjacency graph of the Hamiltonian (on the basis of local spin-orbitals) is depicted. The star between the rings represents a magnetic monopole which produces a central magnetic field with equal  fluxes in the up and down-spin sublattices; however the field has opposite chiralities for opposite spins, and thus represents the effects of the spin-orbit interaction. The vertical bonds represent the effects of the in-plane field $B_{\shortparallel}$ acting on the spin degrees of freedom.
The alternating colors again visualize the bipartite property of the adjacency graph in the spin-orbital basis, which is used for proving the absence of charge current.  }
\label{asymmetric}
\end{figure}


 The time dependent
external magnetic field $\overrightarrow{B}$ has a component
$B_{\shortparallel}(t)$ in the plane of the ring and a component
$B_{\lfloor}(t)$ which produces a flux $\phi(t)=B_{\lfloor} S.$
Let me  write the Hamiltonian for the spin current generation that for various choices of the parameters leads to 3 different operating schemes.
 Each bond in the ring is modified by the field in that it acquires a phase. In the time-dependent case, different distributions of the phase having the same sum over the sides of the ring are not equivalent in principle. Here for simplicity I assume the symmetric prescription $$t_{ring}\rightarrow t_{ring}\exp[{i\alpha(t)/N_{ring}}].$$

  The dynamics of the spin-current production is given by  the Hamiltonian  of Ref. \cite{cibe2}, namely:
\be H_{prod}=H_{D}+H'_{B}\ee\label{hprod} where $H_{D}$ is the device Hamiltonian and
$H'_{B}$ the in plane magnetic term. Here, \be
H_{D}=H_{wires}+H_{ring}+H_{ring-wires}.\ee The polygonal
ring, with an even number $N_{ring}$ of sides, is represented by
\be\label{normalfield}
H_{ring}=\sum_{i\in ring} H_{ring}(i)
\ee
where, with the identification of ring site $N_{ring}+1$ with site 1, we may write
\beq
& H_{ring}(i)=(t_{ring}\exp\left [{i\frac{\alpha(t)}{N_{ring}}}\right ]
 \sum_{\sigma}\exp[i\sigma\alpha_{SO}]c^{\dagger}_{i+1,\sigma}c_{i,\sigma}&\nonumber\\
 &+h.c.)+\mu_{B}B_{\lfloor} \sigma_{z}(i),& \eeq
 with the Bohr magneton $\mu_B=5.79375 *10^{-5}\frac{eV}{Tesla}$ and $\sigma_{z}(i)=\hat{n}_{i+}-\hat{n}_{i-}$.
Here $\alpha_{SO}$ is a phase due to the spin-orbit interaction\cite{Zvyagin}, and can be of order unity or smaller.   The   magnetic interaction due to $B_{\shortparallel}$  acts exclusively on spin and is:
\be H'_{B}=\mu_B B_{\shortparallel}(t) \sum_{i \in
ring}(c^{\dagger}_{i,\uparrow}c_{i,\downarrow}+
c^{\dagger}_{i,\downarrow}c_{i,\uparrow}   ).\ee\label{bdritto}

The  Hamiltonian for the left and right wires is a standard tight-binding model
 \begin{equation} H_{wires}=H_{L}+H_{R}=t_{h}\sum_{n,\sigma} c^{\dagger}_{n,\sigma}c_{n+1,\sigma}+h.c..
  \end{equation}
 Finally, the ring-wires contacts are modeled in $H_{ring-wires}$ whereby the ring has a nearest neighbor connection
to the leads via a tunneling Hamiltonian with hopping matrix elements $t_{lr}$
 in the obvious way. In the numerical calculations  below I take $t_h=t_{ring}=t_{lr}=1$ eV as a reasonable order-of-magnitude. \\

\subsection{Numerical evolution and calculation of the currents}

The system  with a spin-independent chemical potential $E_F$ is in equilibrium for  $t<0$ in the presence of no field. The natural time unit for this problem is $\tau=\frac{\hbar}{t_{h}}.$  In the  code the Hamiltonian is constant during time slices of $ 0.2 \tau$ and jumps to the next value at the end. In this way the many-electron  Schr\"odinger equation is integrated by a succession of sudden approximations.

Taking the spin quantization axis along z (orthogonal to the plane of the ring), the number current operator may be written:
\be
J_{m,\sigma}=\frac{i t_{h}}{\hbar}(c_{m+1,\sigma}^{\dagger}c_{m,\sigma}-c_{m,\sigma}^{\dagger}c_{m+1,\sigma});
\ee
 I calculate the time-dependent number current  by my own partition-free approach\cite{cini80}, namely,
 \be J_{n,\sigma} (t)=-2
\frac{t_h}{h} Im ( G^{<}_{n,\sigma,n-1,\sigma})\ee\label{z} where, in terms of the retarded function $g^{r}_{i,j},$\be
G^{<}_{i,j}(t)=\sum_q n^{0}_q g^{r}_{i,q}(t,0)g^{r
*}_{j,q}(t,0).
\end{equation}
where $q$ runs over the ground state spin-orbitals for $B=0$ and $n^{0}_q$ is the Fermi function. More generally,we need the  current $J_{m}(\overrightarrow{n})$ at site $m$ with the spin quantization axis along $\overrightarrow{n}=(\sin(\theta),0,\cos(\theta)),$in the xz plane ($\varphi=0$). The up-spin  creation operator becomes $$c^{\dagger}_{+,\overrightarrow{n}}=\cos(\frac{\theta}{2})c^{\dagger}_{+}+\sin(\frac{\theta}{2})c^{\dagger}_{-}$$ and the down-spin operator becomes $$c^{\dagger}_{-,\overrightarrow{n}}=-\sin(\frac{\theta}{2})c^{\dagger}_{+}+\cos(\frac{\theta}{2})c^{\dagger}_{-}.$$  Therefore:
 \be J_{m}(\overrightarrow{n})=\cos(\frac{\theta}{2})^2 J_{m,+}+\sin(\frac{\theta}{2})^2J_{m,-}+\sin(\frac{\theta}{2})\cos(\frac{\theta}{2})J^{sf}_{m,+}
 \ee
 where  the spin-flip number current at site $m$ is:
 \be  J^{sf}_{m}=\frac{i t_{h}}{\hbar}\sum_{\sigma}(c^{\dagger}_{m+1,\sigma} c_{m,-\sigma}-c^{\dagger}_{m,\sigma} c_{m+1,-\sigma} ).
 \ee

 Consequently, we are interested in the  the spin current at site $m$  polarized along $\overrightarrow{n}$
 \be
 J^{spin}_{m}(\overrightarrow{n})= J_{m}(\overrightarrow{n})- J_{m}(-\overrightarrow{n})
 \ee
 given in terms of Equation (8)  by
  \be
 J^{spin}_{m}(\overrightarrow{n})= (J_{m,+}- J_{m,-})\cos(\theta)+J^{sf}_{m}\sin(\theta).
 \ee

  My codes calculate number currents taking $t_h =1.$ If this is interpreted to mean that $t_h =1 $eV, which corresponds to the frequency $2.42 *10^{14}$ $s^{-1},$ a current $J=1$ from the code means $2.42 *10^{14}$ electrons per second, which corresponds to a charge current of  $3.87*10^{-5}$ Ampere.
   We also need a characteristic magnetic field.
  Recalling that $\phi_{0}=\frac{h}{e}\sim 4.134 10^{-15}$ in MKSA units, we introduce  the magnetic field $B_{\phi}$ such $B_{\phi}S(N)=\phi_{0}$ where $S(N)=\frac{N a^{2}}{4 \tan(\frac{\pi}{N}}$ is the ring area. Thus,
  $B_{\phi}=\frac{1.65 10^6}{a^2}\frac{\tan(\frac{\pi}{N})}{N}Tesla$, with $a$ in Angstrom.

\section{Magnetic interactions of spinless models}

\begin{figure}[]
\includegraphics*[width=.35\textwidth]{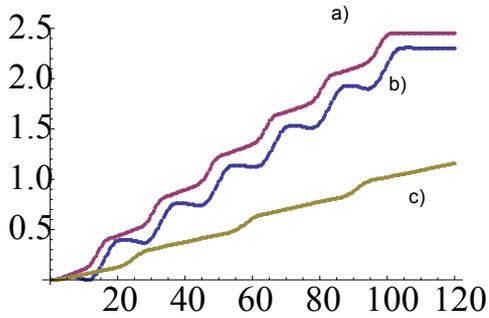}
\caption{ Charge pumped by a laterally connected 6-site ring excited by a time-dependent magnetic flux which grows linearly in time. In this example, spin is not considered.
 Time is measured in units of $\tau=\frac{h}{t_h}$.In all cases,  there is no charging of the ring but charge is transferred from left to right. In a)the system is half filled ($E_{F}=0$ ) and 6 fluxons are inserted in the (0, 100 $\tau $) and each fluxon produces a stair in the pumped charge curve; b)the same but with $E_{F}=0.5 t_{h}$. In   c)$E_{F}=0$  as in a) but the rate of insertion of the flux is reduced, as one can judge from the wider stairs. One can also see a slight reduction of the hight  of the stairs. }
\label{spinless}
\end{figure}

The limiting case  in which we neglect spin in the Hamiltonian  (therefore  $\alpha_{SO}=0$, and  $\mu_{B}=0$ in Equations (5,6)) is already very remarkable\cite{ciniperfetto}.    One method for pumping current while leaving the ring neutral  is  based on the introduction of integer numbers of fluxons, another method consisting in connecting the ring to a junction.  As a direct consequence of the above mentioned nonlinearity of  the magnetic moment, one can achieve, by employing suitable flux protocols, single-parameter nonadiabatic pumping, where an arbitrary amount of charge can be transferred from one side to the other, a phenomenon which, for a linear system, would be readily ruled out by the Brower theorem\cite{brower}. In this way, a current is excited in the wires and charge is pumped in the external circuit without changing the neutrality of the ring. The total energy of the ring after inserting the flux is the same. This happens because inserting an integer number of fluxons is a gauge and the process is slow.  In Figure \ref{spinless} I exemplify the results for a hexagonal ring, showing the charge transferred from a wire to the other as a function of time. Curve a) is obtained at half filling and shows that each fluxon shifts the same amount of charge, producing a staircase. Curve b) is taken at a higher filling, and is not very different. Curve c) is again at half filling, but the stairs are wider because the flux is inserted  more slowly. Moreover the stairs are slightly lower. The amount of charge transferred for each cycle decreases, because the process  is not adiabatic. In the adiabatic limit there is no pumping, as shown in a beautiful analysis by Avron, Raveh and Zur\cite{adiabatic}.

The present treatment neglects interactions. However, the above phenomena were studied within  the Luttinger liquid\cite{ll1} approach. Within the pumping context, a distinct place was attributed to the pumping properties of a Luttinger liquid\cite{ll2, ll3, ll4} and, in particular,it was found that the pumping phenomena persist unhindered in the presence of the interactions\cite{ll5}\\

\subsection{Rotating field or rotating ring}
In the presence of spin it is no longer true that inserting an integer number of fluxons is a gauge; this complicates matters but also produces interesting possibilities. The magnetic moment of the laterally connected bond can pump a partially spin- polarized current even in the absence of a spin-orbit coupling . An example in which a rotating ring produces a partially polarized current is shown in Figure 3. For more data in these arrangements see Ref.(\cite{cibe2}).  When a charge current is present the storage of the spin polarization is complicated by the problem of charging and the theoretical analysis cannot avoid to deal with correlation.  The present paper is devoted to the next arrangement   in which purely spin currents are pumped.

\begin{figure}[]\label{icona}
\includegraphics*[width=.35\textwidth]{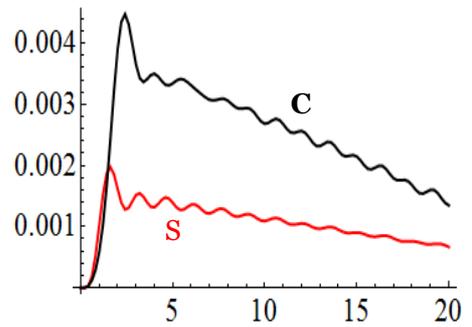}
\caption{Pumping from a 50-site ring laterally connected to wires and rotating with $\omega=\frac{0.004}{\tau}$ in a magnetic field $B=578$ Tesla if $a=1$ Angstrom. Spin current polarized in the direction of the magnetic field (marked S) and charge current (marked C) at the first bond of the right wire. The Fermi energy is $E_{F}=0$ (half filling), but at  $E_{F}=\pm 0.5$ the charge and spin currents are similar to those shown here. No spin-orbit interaction is included in this example.
}
\end{figure}
\subsubsection{Field in plane of ring: pure spin current}
\begin{figure}[]\label{icona}
\begin{center}
\includegraphics*[width=.35\textwidth]{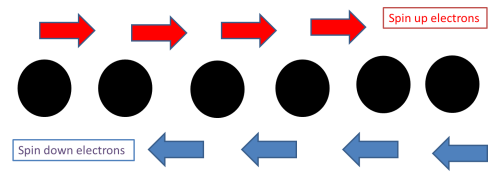}
\caption{A pure spin current does not shift charges but only polarization.It is a novel, fascinating object to study.
 }
\end{center}
\end{figure}
  \begin{figure}[]
\includegraphics*[width=.35\textwidth]{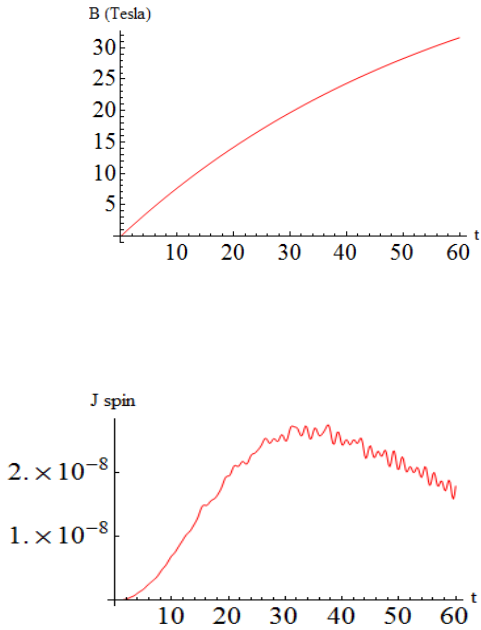}
\caption{Numerical results  of the  model  with B in the plane of the ring  (with 30 sites)  and 100 sites in each lead.$B(t)=B_{max}(1-\exp{-{t\over t_{0}}}), $  with $t_{0}=60$ (the time unit is $\frac{\hbar}{t_{h}}.$) The currents are excited by the  field (upper panel).
The spin current (lower panel) is the same in the two wires, the electron number density is constant everywhere.}
\end{figure}
Now I consider  a geometry such that both spin directions are treated on the same footing. This is obtained by setting the magnetic field in the plane of the ring. In the master Hamiltonian $H_{prod}$ we set $B_{\lfloor}(t)=0$ and $\alpha(t)=0$, since in such an arrangement there no flux piercing the ring. The spin symmetry is broken by $\alpha_{SO}$  which provides  a driving force.   For any time dependence of $V(t)=\mu_{B}B_{\shortparallel}(t)$ the charge current vanishes identically at half filling and a pure spin current (Figure 4) obtains.
 The discrete spin-space symmetry of the model is $P\Sigma$ where $P$ is the parity and $\Sigma$ is the spin reflection. A good analytic understanding of the purely spin current and of its non-adiabatic character was achieved in \cite{cibe1}. Since the adjacency graph for the Hamiltonian in the spin-orbital basis (Fig.1 b) is bipartite for even $N_{ring}$, the absence of a charge current can be deduced from the invariance of the problem under a canonical transformation that exchanges electrons with holes, spin up with spin  down and changes sign to one of the sublattices. The importance of even-odd effects in the physics of quantum rings has been stressed elsewhere\cite{bouncing}

 Numerical results (not shown here) obtained with an odd $N_{ring}$ show a partial spin polarization, in line with the fact that they do not correspond to bipartite lattices. The extra atom produces a charge current that becomes small when the ring gets large. Below we consider even $N_{ring}$.   It was shown\cite{cibe1} that finite temperatures do not change significantly the results up to $K_{B}T \sim 0.025$ eV.  Instead, the results are sensitive to the filling, but for   concentrations  of the order of 0.51   one  gets a spin current with a  small charge current while the ring gets charged.
The  results of Fig.5  were  computed for the full model according to Equation (8).

 The order of magnitude of the spin current for  a short rectangular  spike $V(t)=V\theta(t)\theta(\tau-t)$  with $V(t)=\mu_{B} B_{\shortparallel}(t)$ was found to be
 with the result that   \be J_{s}(\alpha)\sim \frac{t_{h}}{\hbar}\frac{\sin(\alpha)}{2\pi}(\frac{V}{t_{h}})^2,\ee where $ \alpha\sim \alpha_{SO}.$

  As another example I report in Figure 6  the response to an oscillatory field. this is clearly not adiabatic. We see that the current response is far from adiabatic since  the current does not follow the same pattern as the field.

\begin{figure}[]\label{sinusoidal}
\includegraphics*[width=.35\textwidth]{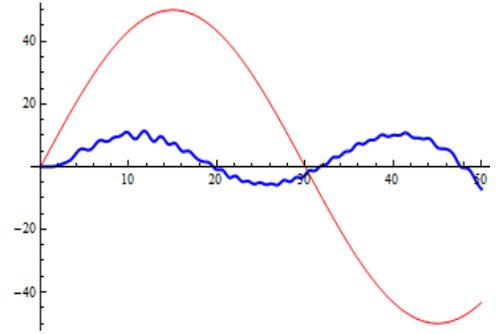}
\caption{Spin current response  of a hexagonal ring to a sinusoidal field $B=B_{max}Sin(\frac{2\pi t}{3 \tau}) $ with $B_{max}=50$ Tesla. Times are measured in units of $\tau$. The spin current is multiplied by $10^8$.}
\end{figure}

 \section{Storage and delayed release of the spin current}
 The possibility of exciting a pure spin current suggests that a magnetization transfer without charging of the kind outlined in the Introduction should  be feasible using a laterally connected ring and a magnetic field in the plane of the ring. It should be possible to store and possibly concentrate the spin current in the form of a long-lived static polarization of an electron gas for later use. Actually the polarization would be permanent in the absence of the spin-orbit interaction. In low atomic number materials the polarization  could be long lived  enough to make interesting experiments on it and maybe technological applications.\\

  To this end, I wish to present the   computer experiment outlined in Figure 7.
 Let the total Hamiltonian bocome:
   \be\label{pila}H=H_{prod}+H_{ext}+H_{L,store}(t)+H_{R,store}(t).\ee
   This is the sum of the above described  $H_{prod}$ plus the terms describing the left and right storage units $H_{L,store}(t), H_{R,store}(t)$ and the external wire $H_{ext}$ of the same form as Equation (6) connecting the storage units. Here $H_{L,store}(t), H_{R,store}(t)$ also contain the connections to the external wire and to the respective wires to the ring, and these depend on time according to the pattern of Figure 7. One must choose $H_{L,store}(t), H_{R,store}(t)$ such that the whole graph is bipartite, and the storage units will be half filled initially and for all times.
    In Figure 7a), the whole system is in equilibrium in the absence of external fields. Then the direct connection is removed and in 7b) the external field in the plane of the ring excites a spin current that will tend to polarize the 8 site  cubes in opposite way. In 7c) the cubes are isolated and the spin current is stored in them as magnetization. Finally,  the direct connection is established in 7d) and a spin current is excited.

 \subsubsection{Storage cubes}

I performed  a numerical experiment in which both storage units where 8-atom cubes.
   The field along the $x$ axis is taken to be $B(t)=B_{max}\sin(\frac{\pi t}{125})$ where  $B_{max}=\frac{B_{\phi}}{100}\sim 58$ Tesla  for $a=10^{-10}$m and a 60 atom ring. The spin-orbit constant is taken to be $\alpha_{SO}=1.$ The leads from the ring to the cubes are 70 sites long while the two cubes are at direct contact with each other (that is, $H_{ext}$ does not exist and there is direct hopping between the polarization reservoirs.). A pure spin current flows in the wires from the ring to the cubes, and its polarization axis is somewhere in the x-z plane; it is more intense when computed with the spin axis along x than along z.  The contacts to the ring are broken at $t_{1}=60 \tau$ while the hopping between cubes is allowed after $t_{2}=70 \tau$; here I recall that  times are measured in units of $\tau=\frac{\hbar}{t_{h}}.$

   In Figure 8 the populations of the two cubes are shown, and it is obvious that they are opposite for up and down spins and also opposite for the two cubes. this realizes the polarization transfer without charging, and is quite substantial. The populations start from half filling and begin to evolve after a time 40 which is needed for the perturbation to cover the 70 site distance from the ring. The speed of the disturbance is about 2 sites in time $\tau$. Then the left cube receives a up polarization and the right ring polarization is opposite. \\

    While this behavior had been expected in the design of the thought experiment, there are surprises, too.  After a (quite large) maximum polarization is achieved, most of it goes away in a rather short time. One could engineer the timing of the process and the size of the storage units if the aim is to maximize the polarization. From $t_{1}$ to $t_{2}$ the rings are isolated and polarization is constant, as it should. The cubes remain strictly neutral like the rest of the system. At $t_{2}$  the cubes are set in direct  contact(in this example $H_{ext}$ is replaced by a direct hopping between the cubes)  and the populations start to oscillate.
     In figure 9 one can observe the immediate, rapidly oscillating purely spin current which is obtained in this way. These oscillations are interesting by their own right and it is likely  that one can  engineer the process, varying the number of sites in the storage units and many available parameters, in order to modify both frequency and amplitude of the oscillations.  The oscillations are complex and fast, and will require further investigations. It should be interesting to observe the spectrum and polarization of electromagnetic waves emitted by these oscillating magnetic currents.
\subsubsection{Storage rings}
   In the next computer experiment the storage cubes are replaced by 4-atom storage rings; the connection $H_{ext}$  between them at the beginning and at the end of the experiment is obtained with a wire composed by $N_{exter}=200$ sites, with hopping $t_{h},$ while the ring is  $N_{ring}=30$  sites large  (See Figure 10). The field is supposed to grow linearly   up to $t=50 \tau$. The spin current in Figure 10 b) is largely reduced, compared to the previous figure, and this is due to the smaller object being magnetized. In Figure  12 c) the length of the wire is shortened, and this produces a shorter delay before the spin current arrives. Moreover the wire connecting the storage rings now has matrix elements $\frac{t_{h}}{3}$ and this produces a  decrease of the overall intensity of the spin current and a slowing down of the oscillations.

\begin{figure}[]\label{storage}
\includegraphics*[width=.35\textwidth]{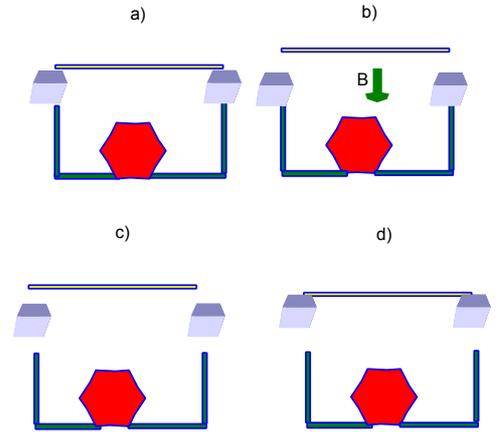}
\caption{The 4 phases of a thought experiment in which a pure spin current is stored in two initially unpolarized cubes and released on command. In a) the storage cubes are linked to a laterally connected ring and directly among themselves. In phase b) the cube-cube connection is removed and a magnetic field produces a spin current. In c) the oppositely polarized cubes are isolated. In d) the cube-cube connection is established and a spin current flows.}
\label{ringgen}
\end{figure}

\begin{figure}[]\label{populations}
\begin{center}
\includegraphics*[width=.35\textwidth]{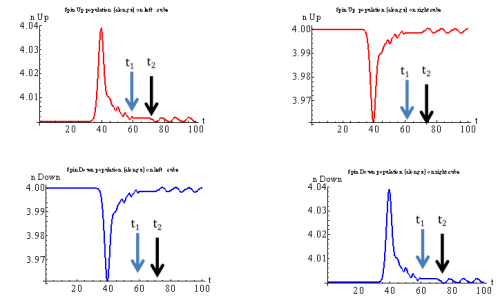}
\caption{Electron populations on the two cubes during the thought experiment. Top left: spin up (along z) on left cube. Top Right:spin up on right cube. Bottom left:spin down on left cube. Bottom right:spin down on right cube. The arrows mark the disconnection from the ring at $t=t_{1}$ and the direct reconnection of the cubes at $t=t_{2}$. If the spin polarization axis is taken parallel to the magnetic field, the trend is similar, but the magnitude of the polarization is about twice as large. }
\end{center}
\label{ringgen}
\end{figure}
\begin{figure}[]\label{Jcubecube}
\begin{center}
\includegraphics*[width=.35\textwidth]{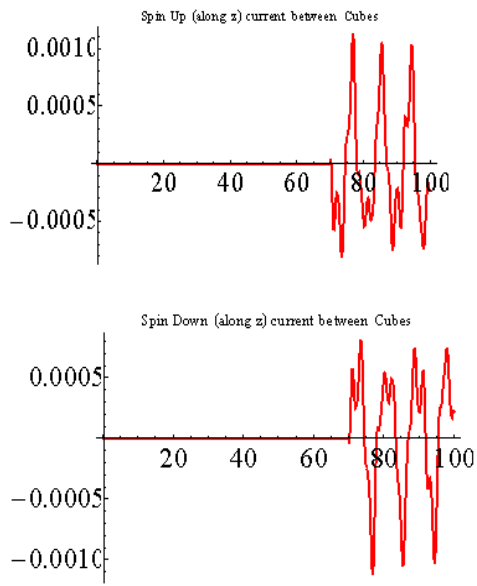}
\caption{The current in the  bond connecting the cubes during the thought experiment. Top:spin up Bottom:spin down.}
\end{center}
\label{ringgen}
\end{figure}
 \begin{figure}[]\label{storagerings}
\begin{center}
\includegraphics*[width=.35\textwidth]{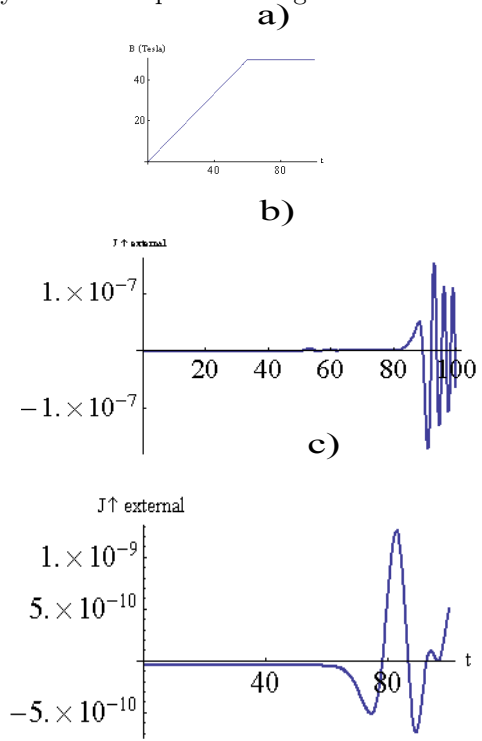}
\caption{A thought experiment in which the storage cubes are replaced by storage rings. a): time dependent magnetic field. b)spin-up current in the 200 atoms long  external (storage ring-storage ring)connection, where the hopping matrix element is $t_{h}$; the leads from the 30 atom central ring ro the storage rings are 50 sites long. The down spin current is opposite.  c) spin-up current in the 100 atoms long  external (storage ring-storage ring)connection, where the hopping matrix element is $\frac{t_{h}}{3}$; the leads from the 30 atom central ring ro the storage rings are 50 sites long. The down spin current is opposite.}
\end{center}
\label{ringgen}
\end{figure}

\section{Conclusions}
I presented the theoretical  study  of  tight-binding model devices consisting of a ring laterally connected to a wire and designed to produce spin polarized currents. Using rotating magnetic fields or letting the ring rotate around the wire one can produce a strong spin polarization even without using the spin-orbit interaction, and this operates in a wide range of fillings. In the case of fast and large rotating rings the centrifugal force must be included and this work is under way.\\

 In another experiment at half filling  a tangent time-dependent magnetic field in the plane of the (fixed) ring can be used to pump a purely spin current, excited by the the spin-orbit interaction in the ring.  This behavior is understood analytically and is found  to be robust with respect to temperature and small deviations from half filling. Above, I presented new numerical results illustrating the above arrangements and the relation to previous findings with spinless models that show charge pumping.

 However the main results of the present paper concern possible schemes to store, gather,and then release magnetization. Suitable  reservoirs or storage units  have been shown to  work with spin currents in analogy with condensers for the usual charge currents. The polarization can be stored and then used on command to produce spin currents in a wire. These currents show oscillations while the storage units exchange their polarizations  and the intensity and amplitude of the oscillations can be modified by changing the conductivity of the ling between the polarized units.

 The present model neglects electron-electron interactions, but  it is   physically reasonable  that adding to the Hamiltonian a correlation term like $U(\hat{n}_{\uparrow}+\hat{n}_{\downarrow}-1)^2$ would tend to reinforce the charge confining  effects described here; in the Hartree approximation, however, it would change nothing since  its average at half filling  vanishes strictly during the evolution of the system.\\

 The above theoretical effort leads to  several intriguing possibilities from the viewpoint of basic research and possible applications and I hope that this will stimulate  experimentalists  to work on the magnetic properties of laterally connected quantum rings.


\end{document}